\documentclass[aps,jcp,onecolumn,preprint,floatfix]{revtex4-2}

\usepackage{graphics} 
\usepackage{color} 
\usepackage{bm} 
\usepackage{bbm}
\usepackage[dvipsnames]{xcolor,colortbl}
\usepackage{amsfonts}
\usepackage[normalem]{ulem}  
\setcitestyle{super}

\definecolor{DarkYellow}{RGB}{80, 80, 0}

\newcommand{\rr}{{\bf r}}
\newcommand{\dr}{{\textrm d}{\bf r}}

\usepackage{amsmath,amssymb,graphicx}
\usepackage{amsbsy}
\usepackage{latexsym}
\usepackage{xcolor}
\usepackage{graphicx}
\usepackage{psfrag}
\usepackage[normalem]{ulem}
\usepackage{bm}
\usepackage{hyperref}
\usepackage[SF,sf]{subfigure}
\usepackage[nottoc]{tocbibind}
\hypersetup{colorlinks=true, linkcolor=red, citecolor=blue, urlcolor=blue}
\usepackage{float}
\usepackage{multirow}
\usepackage{caption}
\begin{document} 

\title{Radial Distribution Function in a Two Dimensional Core-Shoulder Particle System}

\author{Michael Wassermair$^{1,2,3}$}
\email{Michael.Wassermair@ist.ac.at}
\author{Gerhard Kahl$^{1}$}
\email{Gerhard.Kahl@tuwien.ac.at}
\author{Andrew J.~Archer$^{2}$} 
\email{A.J.Archer@lboro.ac.uk}
\author{Roland Roth$^{4}$}
\email{corresponding author; Roland.Roth@uni-tuebingen.de}
\affiliation{$^1$Institut f\"ur Theoretische Physik, TU Wien, Wiedner Hauptstra{\ss}e 8-10, A-1040 Vienna, Austria}
\affiliation{$^2$Department of Mathematical Sciences and Interdisciplinary Centre for Mathematical Modelling, Loughborough University, Loughborough LE11 3TU, United Kingdom}
\affiliation{$^3$Institute of Science and Technology Austria, A-3400 Klosterneuburg, Austria}
\affiliation{$^4$Institute for Theoretical Physics, University of T\"ubingen, D-72076 T\"ubingen, Germany}

\pacs{}
\keywords{~}

\date{\today}

\begin{abstract}
An important quantity in liquid state theory is the radial distribution function $g(r)$.
It can be calculated within the framework of classical density functional theory in two very distinct ways. In the test-particle route, one fixes a single fluid particle, turning it into an external potential in which the inhomogeneous structure of the fluid is calculated by minimising the functional.
The second route to $g(r)$ in density functional theory employs the Ornstein-Zernike equation and the pair direct correlation function, that can be obtained from the second functional derivatives of the excess (over the ideal gas) free energy functional.
Since typically an approximate excess free energy functional is employed, the test-particle route, which requires only one functional derivative, is more accurate than the Ornstein-Zernike route. Here we study a two dimensional core-shoulder particle system and find that in some circumstances the results from the Ornstein-Zernike route can be comparable in accuracy to the test-particle results for $r>\sigma$, the core diameter. We also examine in detail the asymptotic $r\to\infty$ decay of $g(r)$, finding a variety of possible decay wavelengths at different state points and state points where there is a crossover from one wavelength to a very different one. This behaviour is a signature pointing to the rich phase behaviour of the incipient solid phases.
\end{abstract}

\maketitle

\eject

\section{Introduction}

The radial distribution function (RDF) $g(r)$ of a simple fluid characterises the local structure and correlations
between particles and is an important quantity in liquid state theory \cite{hansen13}. It establishes a bridge between the microscopic
description of a fluid and macroscopic quantities such as the internal energy, the pressure or the static structure factor of the 
fluid (see, e.g., Ref.~\citenum{hansen13}). The function $g(r)$
measures the likelihood of finding a particle at distance $r$ from a {\em test}-particle, relative to that in an
ideal gas.
For small separations $r$ the structure of $g(r)$ is strongly influenced by the particle-particle interaction. For fluids
with a hard core of diameter $\sigma=2 R$, like the fluid we study here, $g(r)$ displays a so-called correlation hole, which
implies $g(r\leq \sigma) \equiv 0$. The correlation hole is followed by the first correlation shell of roughly one particle diameter width, 
that contains information about the nearest neighbours of the test particle, followed by the second correlation shell and so on.
Eventually for large distance $r\to \infty$ particles are randomly placed relative to the test particle, like in the case of an
ideal gas, and so as $r\to\infty$, $g(r)\to 1$.
For short-ranged potentials, like the hard core shoulder potential employed in the present study, the decay of $g(r)$ is exponential and
the decay length is the correlation length \cite{hansen13,Evans1994}.
Nevertheless, the form of the decay of $g(r)$ can be very informative and for the system we investigate here can exhibit a rich crossover behaviour from oscillatory decay with one wavelength, to oscillatory decay with a very different wavelength.
Understanding the origin of these different length-scales (wavelengths) is illuminating, because it gives hints towards the possible crystal structures that the system exhibits when the liquid freezes\cite{wassermair2024fingerprints, Wassermair2026}.
Because of the importance of $g(r)$ and the quantities that can be derived from it, there are many different approaches
to compute this function, including classical density functional theory (DFT) \cite{Mermin1965,Evans1979,hansen13}, integral equation theory that
combines the Ornstein-Zernike (OZ) equation \cite{OZ1914} with a closure relation \cite{hansen13}, and computer simulations \cite{Metropolis1953}.

In this manuscript we employ DFT to study $g(r)$ for a hard core 
square-shoulder 
fluid in two dimensions (2D) with an interaction potential that is given by
\begin{equation}\label{eq:pot}
	\phi(r)=\begin{cases}
		\infty & r \leq \sigma ~,\\
		\epsilon &\sigma < r <\lambda\sigma ~, \\
		0 &\lambda\sigma \leq r ~, \\
	\end{cases}
\end{equation}
where $\lambda \sigma$ is the range of the (repulsive) square-shoulder interaction with a height $\epsilon>0$. Within DFT 
one can prove \cite{Mermin1965,Evans1979,hansen13} that there exists the grand potential functional $\Omega[\rho(\bf r)]$, that it is 
a functional of the ensemble averaged one-particle density $\rho(\bf r)$, and that it is minimised by the equilibrium density distribution 
$\rho_0(\bf r)$, for which the functional reduces to the grand potential $\Omega=\Omega[\rho_0(\bf r)]$ of the system. This 
can be written in the compact form of the Euler-Lagrange equation \cite{Evans1979}
\begin{equation} 
\label{eq:dOmega_drho}
\left. \frac{\delta\Omega[\rho] }{\delta\rho(\bf r)}\right|_{\rho(\bf r)=\rho_{0}(\bf r)}=0.
\end{equation}

Unfortunately, the mathematical proof of DFT, that guarantees the existence of the functional does not provide insight in how
the functional, especially the part that describes inter-particle interaction, is constructed. Several approximations for hard spheres
in dimension $d=3$ have been found \cite{Tarazona1984,Tarazona1985,Rosenfeld1989}, the most successful being fundamental measure
theory (FMT) \cite{Rosenfeld1989,Roth2010}. FMT follows the structure of the exact DFT for hard rods in $d=1$ 
\cite{Percus1976,Vanderlick1989}, which makes use of the geometrical measures of the particle shapes to account for
the hard core repulsion. FMT functionals \cite{Rosenfeld1989,Roth2002,Ju2002,HansenGoos2006,Lutsko2020} provide a
very accurate description of hard sphere fluids compared to computer simulations. The ideas of FMT have been
successfully transferred  \cite{Rosenfeld1990,Roth2012} and applied to hard disk fluids in $d=2$ \cite{Thorneywork2018}.
Within the framework of DFT the soft (finite) part of the inter-particle interaction in Eq.~(\ref{eq:pot}), is typically treated by
a random phase approximation (RPA) \cite{hansen13,Archer2017}. 

DFT offers two very distinct routes to calculate $g(r)$. The first approach is called the 
test-particle route and was suggested by Percus \cite{Percus1962,Percus1964}, who realised that a particle of the fluid can be fixed 
at e.g.\ the origin of our coordinate system and thereby be made an external potential acting on the rest of the fluid. The external 
potential $V_{\textrm{ext}}(r) = \phi(r)$ is the pair potential of the fluid, which in the present case is radially symmetric. If one 
minimises the density functional to obtain the equilibrium density profile $\rho(r)$ in this external potential, it is related to the 
RDF via $g(r) = \rho(r)/\rho_{\textrm{b}}$, where $\rho_{\textrm{b}}$ is the bulk density far away 
from the test particle. This approach requires one to only take one functional derivative of the {\em approximate} grand potential functional 
$\Omega[\rho]$ in order to solve the Euler-Lagrange  equation, Eq.~(\ref{eq:dOmega_drho}), which can be done numerically using e.g.\ a Picard iteration.

The second approach makes use of the OZ relation, which connects the total correlation function $h(r) = g(r)-1$
with the so-called pair direct correlation function (pDCF) $c^{(2)}(r)$ via \cite{OZ1914,hansen13}
\begin{equation}\label{eq:OZ}
	h(r)=c^{(2)}(r)+\rho_{\textrm{b}} \int c^{(2)}(|{\bf r}-{\bf r'}|) h(r') d {\bf r'}.
\end{equation}
In order to solve this equation an additional closure relation which relates the functions $h(r)$ and $c^{(2)}(r)$ is required. However, within DFT it is possible to calculate the pDCF
$c^{(2)}(r)$ from the excess free energy functional ${F}_{\rm ex}[\rho({\bf r})]$ \cite{Evans1979}
\begin{equation}\label{eq:c2}
	c^{(2)}(r = |{\bf r}-{\bf r'}|)=-\beta\frac{\delta^2{F}_{\rm ex}[\rho({\bf r})]}{\delta\rho({\bf r})\delta\rho({\bf r'})},
\end{equation}
which requires the second functional derivative of the {\em approximate} density functional. 
From a FMT functional it is possible to obtain a closed analytical expression for the pDCF, which
can be employed to solve the OZ equation, i.e.\ the approximate excess free energy functional 
${F}_{\rm ex}[\rho({\bf r})]$ replaces the need for an additional (approximate) closure relation.

The core condition $g(r \le \sigma)\equiv0$ for hard cores is often part of a closure relation \cite{PY1958,hansen13}, but is usually violated 
when a pDCF obtained via Eq.~\eqref{eq:c2} is employed in the OZ equation \cite{Schmidt2000}. This is acceptable as 
long as the RDF outside of the core is accurate.
Within DFT the typical experience is that $g(r)$ 
predicted by the test-particle route is more accurate than the one obtained via the pDCF.
The explanation or 
rationalisation for this comes from the fact that a computation that requires only one functional derivative of an approximate functional should
be more reliable than one that requires two functional derivatives. 
Another rationalisation is the one discussed in \cite{Archer2017}, based on the observation that the test-particle Euler-Lagrange equation can be written in the form of the OZ equation, that turns out to have a hybrid closure relation that is better than one might expect from just considering Eq.~\eqref{eq:c2}. 
We discuss this issue in more detail in Appendix A of this paper.
A somewhat surprising outcome of the present work is the observation that for $r>\sigma$ the OZ result is almost as accurate as the test-particle result for sufficiently large values of the shoulder range $\lambda$.
This is valuable, because the OZ expressions are analytic in Fourier space.
This provides the basis for an analysis of the asymptotic decay behaviour of $g(r)$, which turns out to be very rich, exhibiting multiple crossovers in the phase diagram from damped oscillatory decay with one wavelength, to damped oscillatory decay with a very different wavelength.

In this manuscript we employ DFT for a two dimensional (2D) hard core square-shoulder fluid 
\cite{wassermair2024fingerprints,Wassermair2026} in order to predict the RDF $g(r)$ using the test-particle route 
and compare these results to $g(r)$ obtained via the OZ equation together with a pDCF obtained 
from the same density functional. The manuscript is structured as follows. We present the theory in Sec.~\ref{sec:theory}, where we specify the details 
of the density functional employed here, give details about the Monte-Carlo simulation approach used for obtaining benchmark $g(r)$ data and other details relating to how we perform the calculations. In Sec.~\ref{sec:asymptotics} we briefly discuss the theory for determining the $r\to\infty$ asymptotic decay of $g(r)$. In Sec.~\ref{sec:results} we present our results. In 
Sec.~\ref{sec:summary} we make a few concluding remarks. In Appendix A we briefly discuss how the OZ equation in combination with the  random phase approximation (RPA) closure relates to the RPA-DFT Euler-Lagrange equation in the test-particle limit, while in Appendix B we give a few of the technical details relating to our DFT calculations.

\section{Theory} \label{sec:theory}

We consider a 2D one-component fluid with pair interactions between the particles, within the framework of DFT.
The grand potential functional is given by \cite{Evans1979, hansen13}
\begin{equation}\label{DFT2}
	\Omega\left[ \rho \right]=F_{\rm id}[\rho]+
	F_{\rm \textrm{ex}}[\rho] + \int \rho({\bf r}) \left[V_{\textrm{ext}}({\bf r})-\mu\right]d {\bf r},
\end{equation} 
where $\mu$ is the chemical potential and $V_{\textrm{ext}}({\bf r})$ is the external potential. 
The ideal gas contribution to the free energy in two dimensions $F_{\rm id}[\rho] $ is known exactly and is given by
\begin{equation}\label{DFT3}
F_{\rm id}[\rho] = k_{\rm B} T \int \rho({\bf r}) \left[ \ln
	\Lambda^2 \rho ({\bf r}) - 1 \right] d {\bf r},
\end{equation}
with the (irrelevant) thermal de Broglie wavelength $\Lambda$. The excess (over the ideal gas) Helmholtz free energy functional 
$F_{\rm ex}[\rho]$, which contains all the information about the inter-particle interaction, can be decomposed, within perturbation theory, into two contributions 
\begin{equation}\label{eq:split}
    F_{\rm ex}[\rho]=F_\mathrm{c}[\rho]+F_\mathrm{sh}[\rho].
\end{equation}
The first term incorporates the contribution due to the hard core particle interactions and represents the reference system
for the fluid with the interaction given in Eq.~(\ref{eq:pot}). It is approximated within
fundamental measure theory (FMT)
\cite{ Rosenfeld1989,Roth2010, Roth2012,wassermair2024fingerprints}, which makes the ansatz that the excess free energy functional
is a volume integral of an excess free energy density, that is a function of position-dependent weighted densities. For a fluid of hard disks the related state of the
art functional is given by \cite{Roth2012}
\begin{multline}\label{FMT_ex}
F_\mathrm{c}^{\rm FMT}[\rho] = k_{\rm B }T \int \Bigg[ -n_0 ({\bf r}) \ln{(1-n_2 ({\bf r}))}\\
+\frac{1}{4\pi(1-n_2 ({\bf r}))} \left( \frac{19}{12}({\bf n}^{(0)}({\bf r}))^2-\frac{5}{12}{\bf n}^{(1)}({\bf r})\cdot {\bf n}^{(1)}({\bf r})-\frac{7}{6}{\bf n}^{(2)}({\bf r}) \cdot {\bf n}^{(2)}({\bf r})\right) \Bigg] d{\bf r},
\end{multline}
with weighted densities $n_\alpha ({\bf r})$ (with $\alpha = 0,2$) and ${\bf n}^{(m)}({\bf r})$ (with $m = 0, 1, 2$)
that are convolutions of the local density $\rho(\bf r)$ with geometrical weight functions $\omega_\alpha(\bf r)$ and $\omega^{(m)}(\bf r)$. For hard disks one requires two
scalar weighted densities
\begin{equation}
n_\alpha ({\bf r}) = 
\left[ \rho\otimes\omega_\alpha \right] ({\bf r}), \hspace{1cm} \alpha= 0, 2,
\end{equation}
and three tensorial weighted densities
\begin{equation}
{\bf n}^{(m)}({\bf r}) = 
\left[ \rho\otimes{\bf \omega}^{(m)} \right] ({\bf r}), \hspace{1cm} m=0, 1, 2 .
\end{equation}

The weight functions are \cite{Roth2012}
\begin{equation}
\omega_0(r)=\frac{\delta(R-r)}{2\pi R} ~~~~ {\rm and} ~~~~~
\omega_2(r)=\Theta(R-r),
\end{equation}
where $R=\sigma/2$ is the radius of the disks, $\delta(r)$ is the Dirac delta 
distribution and $\Theta(r)$ is the Heaviside step-function. The tensorial weight functions are given by 
\begin{equation}
{\bf \omega}^{(m)}({\bf r})=\delta(R-|{\bf r}|)\underbrace{{\bf \hat{r}}\dots{\bf \hat{r}}}_{\text{$m$-times}}.
\end{equation}
The rank $m$ tensorial weight function arises from taking $m$ tensor products of the unit vector ${\bf \hat{r}}$ with itself.

The second term in Eq.~\eqref{eq:split} is the following RPA approximation \cite{Evans1979, hansen13, Archer2017,wassermair2024fingerprints}
\begin{equation}\label{DFT_RPA_fex}
F_\mathrm{sh}[\rho]=\frac{1}{2}\int\int\rho({\bf r})\rho({\bf r'})\phi_\mathrm{sh}(|{\bf r}-{\bf r'}|) d{\bf r} d{\bf r'},
\end{equation}
where
\begin{equation}\label{eq:pot_shoulder}
	\phi_\mathrm{sh}(r)=\begin{cases}
		\epsilon &0 < r <\lambda\sigma ~, \\
		0 &\lambda\sigma \leq r ~, \\
	\end{cases}
\end{equation} 
is the repulsive shoulder part of the pair potential.
Note that the repulsion has been extended inside the core of the particles. 

Having fully specified the excess free energy functional, we are in the position to compute the pDCF using
Eq.~(\ref{eq:c2}). Since the OZ equation, Eq.~(\ref{eq:OZ}), can most easily be solved in Fourier space to give
\begin{equation} \label{eq:OZk}
\hat{h}(k) = \frac{\hat{c}^{(2)}(k)}{1- \rho~\hat{c}^{(2)}(k)},
\end{equation}
we require the Fourier transform of the pDCF $\hat{c}^{(2)}(k)$. Note that the functional form of the 
excess free energy functional given in \eqref{eq:split} implies that the pDCF, within perturbation theory, can be split into two terms,
which obviously also holds for its Fourier transform
\begin{equation}\label{eq:c_split}
\hat{c}^{(2)}(k) = \hat{c}_\mathrm{c}^{(2)}(k) + \hat{c}_\mathrm{sh}^{(2)}(k).
\end{equation}
The first term, arising from the hard disk core repulsion treated using Eq.~\eqref{FMT_ex}, makes use of 
the structure of FMT. One finds that the core contribution to the pDCF, Eq.~(\ref{eq:c2}) can be written as 
\begin{equation}
c_\mathrm{c}^{(2)}(r=|{\bf r_1}-{\bf r_2}|)  = - \sum_{\alpha,\gamma}
\frac{\partial^2 \Phi}{\partial n_\alpha \partial n_\gamma} \int d
{\bf r}' \omega_\alpha({\bf r}'-{\bf r_1}) \omega_\gamma({\bf
  r}'-{\bf r_2}),
  \label{eq:c2_HD}
\end{equation}
where $\Phi$ is the integrand in Eq.~\eqref{FMT_ex}. Equation~\eqref{eq:c2_HD} can be transformed into Fourier space with the help of the convolution theorem, giving 
\begin{equation}
\hat{c}_\mathrm{c}^{(2)}(k)  = - \sum_{\alpha,\gamma}
\frac{\partial^2 \Phi}{\partial n_\alpha \partial
  n_\gamma}~\hat{\omega}_\alpha(k)~ \hat{\omega}_\gamma(-k).
\end{equation}
The second derivatives of $\Phi$ w.r.t.\ the weighted densities and the
Fourier transforms of the weight functions are known analytically and
hence we arrive at the following analytical expression for $\hat{c}_\mathrm{c}^{(2)}(k)$ \cite{wassermair2024fingerprints, Thorneywork2018}
\begin{multline} \label{c_FMT_hc}
\hat{c}_\mathrm{c}^{(2)}(k)=\frac{\pi}{6(1-\eta)^3k^2}\bigg[-\frac{5}{4}(1-\eta)^2k^2J_0(k/2)^2\\+\left( 4((\eta-20)\eta+7)+\frac{5}{4}(1-\eta)^2k^2\right)J_1(k/2)^2\\+2(\eta-13)(1-\eta)k J_1(k/2)J_0(k/2)\bigg],
\end{multline}
where $J_n(x)$ are Bessel functions of order $n$.
The second term in Eq.~\eqref{eq:c_split} is the contribution from the shoulder to the pDCF, generated by the functional in Eq.~\eqref{DFT_RPA_fex}, and may be written as \cite{wassermair2024fingerprints}
\begin{equation} \label{phi:ss_k}
\hat{c}_\mathrm{sh}^{(2)}(k) = -\beta2\pi\epsilon\lambda\frac{J_1(\lambda k)}{k} .
\end{equation}
With the explicit expression for $\hat{c}^{(2)}(k)$, the total correlation function $h(r)$ can be calculated via an inverse Fourier transform.

As benchmark data for our DFT results we perform grand canonical
Monte-Carlo (GCMC) simulation of the square-shoulder system, to obtain $g(r)$.
The GCMC simulations for the liquids were performed in square boxes of size $125\sigma\times125\sigma$. A total of $20\times10^{9}$ 
GCMC-moves was performed, where in each step particle translation, creation and deletion was attempted with equal probability $\alpha_{\rm t}=\alpha_{\rm c}=\alpha_{\rm d}=\frac{1}{3}$. For the densities considered the simulations contained roughly 2000 - 8000 particles. The RDF is calculated from the positions of the particles, {\bf r}$_i$ via \cite{hansen13}
\begin{equation}\label{ST1}
	g(r)= \frac{1}{\rho_\mathrm{b}}\left\langle \frac{1}{N}\sum_{i=1}^{N}\sum_{j=1,j\neq i}^{N}\delta({\bf r}+ {\bf r}_i-{\bf r}_j) \right\rangle,
\end{equation}
where the brackets denote a grand-canonical average. This ensemble average was performed over 250 configurations, which were
separated by $20\times10^6$ MC-moves along a Markov-chain. 

\section{Asymptotic decay of correlations}
\label{sec:asymptotics}

The properties of $g(r)$ at intermediate values of $r>\sigma$ reveals much about the local packing environment of the particles in a fluid. However, significant understanding of the structure in a fluid can also be understood from inspecting the $r\to\infty$ decay behaviour of $g(r)$.
The general theory for the decay of $g(r)$ was developed by Evans and co-workers, initially for three-dimensional (3D) fluids \cite{evans1993asymptotic, Evans1994, carvalho1994decay} and then more recently for 2D fluids \cite{walters2018structural}. The starting point for the analysis is Eq.~\eqref{eq:OZk}. Taking the inverse Fourier transform in 2D we obtain
\begin{equation} \label{eq:inverse_OZ}
h(r) =\frac{1}{2\pi}\int_0^\infty dk\, k J_0(kr) \frac{\hat{c}^{(2)}(k)}{1- \rho\,\hat{c}^{(2)}(k)},
\end{equation}
where $J_0$ is the zeroth Bessel function of the first kind. In 3D there is an equivalent but somewhat simpler formula that involves an exponential, rather than a Bessel function \cite{evans1993asymptotic, Evans1994, carvalho1994decay}.
The key idea is to evaluate this integral as a contour integral in the complex-$k$ plane.
The contour chosen is typically a semi-circle in the upper half of the complex plane \cite{evans1993asymptotic, Evans1994, carvalho1994decay, walters2018structural}, but other choices are sometime possible \cite{frusawa2026symmetric}.
Evaluating the integral \eqref{eq:inverse_OZ} in this manner transforms it into a sum over residues of the poles of the integrand in the upper half of the complex plane. However, as we show below, the pole structure in the upper half of the complex plane is mirrored in the lower half, so evaluating via a contour around the lower half plane, is equally possible.
The poles arise at points in the complex-$q$ plane where the denominator in the integrand of \eqref{eq:inverse_OZ} is zero, i.e.\ where
\begin{equation}\label{eq:pole_condition}
    [1- \rho\hat{c}^{(2)}(q)]=0.
\end{equation}
Note that henceforth we denote complex wavenumbers with the letter $q$, while we denote real wavenumbers with the letter $k$.
Typically, there are very many (possibly an infinite number) of poles, i.e.\ roots of \eqref{eq:pole_condition}. However, it is the pole(s) $q_n=k_r+ik_i$ with smallest imaginary part $k_i$ which determines the asymptotic decay of $h(r)$, since each pole (together with its complex conjugate pair) contributes a term $\sim\exp(iq_nr)$ to $h(r)$ \cite{evans1993asymptotic, Evans1994, carvalho1994decay, walters2018structural}. 
If the pole is purely imaginary, then the asymptotic $r\to\infty$ decay of $h(r)$ takes the form
\begin{equation}
    h(r)= \frac{A_n\exp(-k_ir)}{\sqrt{r}}+f,
\end{equation}
where ${f}$ denotes terms having a {\em faster} decay, as $r\to\infty$. A complex pole together with its conjugate pair, instead lead to a decay of the form \cite{walters2018structural}
\begin{equation}\label{hr_asym}
    h(r)= \frac{A_n\exp(-k_ir)\cos(k_r r+\theta)}{\sqrt{r}}+f,
\end{equation}
where $\theta$ is a phase shift \cite{evans1993asymptotic, Evans1994, carvalho1994decay, walters2018structural}. There are very similar results in 3D, except in 3D the $\sqrt{r}$ in the denominator is replaced just by $r$ \cite{evans1993asymptotic, Evans1994, carvalho1994decay}. 
When there are two pairs of poles (i.e.\ four poles in total) that have the same value for their imaginary part $k_i$, then there is a crossover from oscillatory decay with one wavelength, to oscillatory decay with another wavelength as on moves through that point in the phase diagram. Only a few one-component systems are known to exhibit such a crossover  \cite{archer2007model, walters2018structural}.
It is much more common in binary mixtures, where there is often a crossover in the decay of the correlation functions $g_{ij}(r)$ as the relative concentrations of the two species $i$ and $j$ are varied, as long as there is a sufficient size difference between the particle size of the two species \cite{archer2001binary, grodon2004decay, grodon2005homogeneous, baumgartl2007experimental, statt2016direct}.
Note also that there can be a crossover from monotonic decay to damped oscillatory decay.
The line in the phase diagram at which this occurs is known as the Fisher-Widom line \cite{fisher1969decay, evans1993asymptotic}.

The poles (and therefore the asymptotic decay) can be determined by solving Eq.~\eqref{eq:pole_condition} for complex $q$. A related quantity is
\begin{equation}\label{eq:S}
    S(q)=\frac{1}{1 - \rho\hat{c}^{(2)}(q)},
\end{equation}
which when evaluated for {\em real} $q=k_r=k$ yields the static structure factor $S(k)$ \cite{hansen13}. Below we present results for the locations of the poles $q_n$ and on the same plots display $\arg S(q)$, which is also illuminating.

As well as the static structure factor $S(k)$, another related and physically relevant quantity is the dispersion relation \cite{Archer2004, archer2012solidification, Wassermair2026}
\begin{equation}\label{eq:omega}
    \omega(k)=-D k^2[1 - \rho\hat{c}^{(2)}(k)],
\end{equation}
where $D$ is the diffusion coefficient.
The dispersion relation determines the linear (small amplitude) growth or decay of density fluctuations in the uniform liquid.
When one considers a small density perturbation of the liquid density of the form $\rho(\mathbf{r},t)=\rho+\delta \rho(\mathbf{r},t)$, one finds that this evolves subsequently over time as the following Fourier sum
\begin{equation}\label{eq:d_rho}
\delta\rho(\mathbf{r},t)=\sum_\mathbf{k}\hat{\rho}_\mathbf{k}e^{i\mathbf{k}\cdot\mathbf{r}+\omega(k)t},
\end{equation}
as long as the Fourier amplitudes $\hat{\rho}_\mathbf{k}$ are small.

Since $\omega(k)$ is a growth/decay rate, the wavenumbers $k=k_u\neq0$ where $\omega(k)$ has a local maximum (i.e.\ where $\partial\omega(k_u)/\partial k=0$) are physically relevant, since when $\omega(k_u)>0$, these are the fastest growing modes.
The locus in the phase diagram where the largest local maximum in $\omega(k)$ at $k=k_u^\mathrm{max}$ has growth rate $\omega(k_u^\mathrm{max})=0$ is referred to here as the linear stability threshold.
To one side of this line in the phase diagram, we have $\omega(k)<0$ for all $k\neq0$ and so the uniform liquid is linearly stable, while on the other side we have $\omega(k)>0$ for $k\approx k_u^\mathrm{max}$ and so the liquid is unstable.
This is where crystalline or quasicrystalline phases are to be expected \cite{archer2012solidification, Wassermair2026}.
Note that this threshold is sometimes referred to as the $\lambda$-line~\cite{ciach2003effect, archer2004soft}. However, to avoid confusion with the shoulder range parameter, here we avoid referring to it this way.
Owing to the close connections between $\omega(k)$ and $S(q)$ -- see Eq.~\eqref{eq:pole_condition} -- there are close connections between the asymptotic decay of $h(r)$ and shape of $\omega(k)$, that we elucidate further below.

\section{Results} \label{sec:results}

In the following we make use of some reduced units. We set the hard disk diameter $\sigma=1$ as the unit of length and hence measure
all other lengths in units of $\sigma$. The temperature $T$ of the system sets an energy scale $k_{\rm B} T$, where $k_{\rm B}$ is
the Boltzmann constant. In the present system, the only other energy scale that can be compared to $k_{\rm B} T$ is the shoulder
height $\epsilon$ in Eq.~(\ref{eq:pot}), which enters our calculation as the dimensionless quantity $\beta \epsilon$, where 
$\beta = 1/(k_{\rm B} T)$, as usual. To change the temperature in our system we {\em effectively} set $\beta \equiv 1$ and use,
following our previous convention \cite{wassermair2024fingerprints}, $T=1/\epsilon$.
In this study our main interest is in the fluid phase.
However, we do also present results for some quantities at state points where the solid phases arise in order to (i) show the contrast with properties in the liquid and (ii) to connect to the work in Ref.~\cite{Wassermair2026}, on the solid phases.


\begin{figure}
\centering
\includegraphics[width=1.0\textwidth]{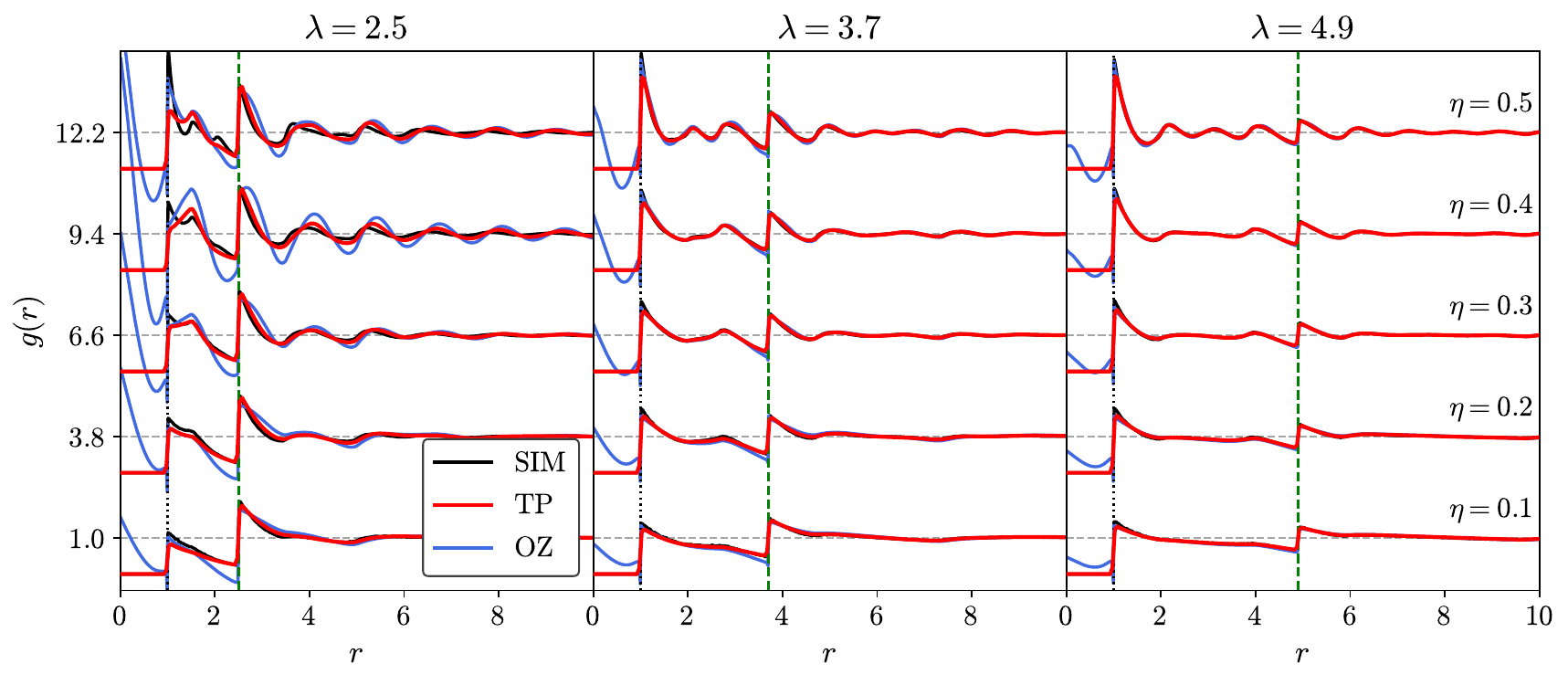}
\caption{RDFs $g(r)$ as functions of $r$ for the core-shoulder fluid, calculated for three different values of the shoulder range $\lambda$ (increasing from left to right, as labelled) and for five different values of packing fraction $\eta$ (decreasing from top to bottom, as labelled). For better visibility, the curves are shifted vertically, with a 2.8 unit shift between each set of curves. The simulation (SIM) results are the black lines, the test-particle (TP) results are the red lines and data obtained via the Ornstein-Zernike (OZ) approach outlined in the text are the blue lines. The temperatures of the states are located at $T=1.25\times T^{\lambda}_{\text{max}}$, where $T^{\lambda}_{\text{max}}$ are the temperatures of the maxima of the respective linear stability threshold lines. The corresponding temperature values are:  $T=0.4748$ ($\lambda=2.5$), $T=0.8827$ ($\lambda=3.7$), and $T=1.4618$ ($\lambda=4.9$).}
\label{fig:1}
\end{figure}

We start by presenting results for $g(r)$ for three different values of the interaction range, $\lambda=2.5$, 3.7 and 4.9. These rather large values of the square-shoulder range give rise to interesting crystalline and quasicrystalline behaviour at low temperatures \cite{Wassermair2026}. First, we discuss
the RDF $g(r)$ obtained using the test-particle approach (red curves in Fig.~\ref{fig:1}). These are calculated by fixing one 
particle at the origin and turning it into an external potential for the rest of the system. The external potential is set to be equal to the pair potential, $V_{\textrm{ext}}(\rr) = \phi(r)$.
It might seem best to make use of the radial symmetry of
the problem and reduce the resulting DFT calculation to an effective one-dimensional (1D) problem. However, within FMT one reason
for its good performance stems from the fact that the weighted densities are in $\rr$-space convolutions of the local density with geometrical weight
functions, which can be evaluated fast and accurately using fast Fourier transforms (FFTs). If  we reduce the dimensionality of the
problem from 2D to effective 1D, weighted densities lose the property of being convolution products and the calculation becomes
far less efficient and the computation time increases by orders of magnitude. For this reason we stay with a full 2D system 
to perform our calculations, using uniformly discretized Cartesian coordinates, which do not allow for an exact representation of the hard core of the external
potential. This problem can be mitigated by averaging the resulting density profile $\rho({\bf r})=\rho(r,\varphi)$ over the angle $\varphi$ to obtain
\begin{equation}\label{eq:angav}
g(r)=\frac{1}{2\pi}\int_0^{2\pi}\frac{\rho(r,\varphi)}{\rho_\mathrm{b}}d\varphi,
\end{equation}
where $\varphi$ is the polar angle of {\bf r}.
A few further details regarding our DFT calculations are given in Appendix B.
While there are alternative numerical approaches, that allow for a computationally efficient reduction of the 2D to an
{\em effective} 1D problem like quasi-spectral methods \cite{Nold2017} or employing a logarithmic grid 
\cite{Hamilton1999,Oettel2009}, we decided to stick to the 2D implementation, because the results presented
here are part of a larger project \cite{wassermair2024fingerprints,Wassermair2026}.

In Fig.~\ref{fig:1} we show our DFT result for the test-particle $g(r)$, calculated using FMT (red lines) for five different values of the bulk packing fraction $\eta=\pi \rho_\mathrm{b}/4$, where we have made use of $\sigma=1$, for a fixed shoulder range 
$\lambda=2.5$, 3.7 and 4.9. The calculations are performed at a fixed reduced temperature $T=1.25\times T^{\lambda}_{\text{max}}$, where $T^{\lambda}_{\text{max}}$ is the temperature of the maxima of the linear stability threshold line \cite{Wassermair2026} for the given value of the shoulder range $\lambda$.
The respective temperature values are $T=0.4748$ for $\lambda=2.5$, $T=0.8827$ for $\lambda=3.7$, and $T=1.4618$ for $\lambda=4.9$.
Due to the hard core of the fixed test-particle at the origin, the RDF $g(r)$ is enforced to vanish 
for $r<1$.
At $r=1$ the RDF makes a jump from zero inside the core to its contact value.
For $r>1$ the RDF displays an oscillatory behaviour due to packing effects.
Note also that at $r=\lambda$ the interaction 
potential $\phi(r)$ and hence the external potential in the test-particle geometry jumps from $\epsilon>0$ to zero and as a result
the RDF also changes discontinuously at $r=\lambda$.

We find that $g(r)$ displays a complex, oscillatory behaviour, which becomes more pronounced as the 
packing fraction $\eta$ increases. This feature is well captured by the test-particle results (red lines), which is not too surprising because FMT accounts for short ranged correlations due to hard cores. In order to test the quality of the test-particle results within DFT we compare them to GCMC data, which are shown as black lines in Fig.~\ref{fig:1}.
We find that the overall behaviour of the simulation-based $g(r)$ is captured well by the test-particle results, albeit with contact values of $g(r)$  and its values at the square shoulder distance $r=\lambda$ a little too small.
The accuracy of the test-particle results improve with increasing shoulder range $\lambda$.
It is worth mentioning that the structure of $g(r)$ for hard disks and for hard disk mixtures can be accounted for by FMT test-particle with much higher precision \cite{Roth2012,Thorneywork2018} compared to the results presented in Fig.~\ref{fig:1}. 

As already mentioned, a second approach to $g(r)$ within DFT makes use of the OZ equation~(\ref{eq:OZ}). 
Instead of a closure relation, we employ the pDCF from DFT, derived from the excess free energy functional via
Eq.~(\ref{eq:c2}).
It is important to realise that for hard disks the OZ route to $g(r)$ is typically significantly less accurate than the test-particle route, while it still predicts reasonable results for the pDCF obtained from FMT.
The results from this OZ route are displayed in Fig.~\ref{fig:1} as blue lines.
The first observation we can make is that the core condition $g(r<1)=0$ 
is violated by the OZ route. This observation has been made before for the OZ route using a pDCF obtained from an approximate excess free-energy functional, see e.g.\ Ref.~\cite{Schmidt2000}. In order to avoid 
such a deficiency, closure relations like the Percus-Yevick closure \cite{PY1958,hansen13},
enforce the core condition. 
However, we also observe in Fig.~\ref{fig:1} that outside the core, the OZ results show a reasonable
agreement with the computer simulations (black lines). Furthermore, the OZ route results for larger $\lambda$, such as $\lambda=3.7$ and 4.9, are comparable in accuracy to those from the test-particle route, agreeing outside of the core surprisingly well with our simulation results. 

These observations are valuable, because the pDCF contribution due to the core can be calculated analytically within FMT in Fourier space, resulting in the expression for $\hat{c}_\mathrm{c}^{(2)}(k)$ in terms of Bessel functions given in Eq.~(\ref{c_FMT_hc}).
The shoulder contribution to the pDCF can also be calculated analytically in Fourier space and again contains a Bessel function; see Eq.~(\ref{phi:ss_k}).
While in odd dimensions ($d=1,3,\dots$) the Fourier transforms of the FMT weight 
functions can be expressed with trigonometric functions, in even dimensions ($d=2,4,\dots$) one finds Bessel functions instead. Once 
the pDCF in Fourier space is given, the OZ equation can be solved using Eq.~(\ref{eq:OZk}) via 
an inverse Fourier transform. In $d=2$ this results in an inverse Hankel transform.

\begin{figure}
\centering
\includegraphics[width=1.0\textwidth]{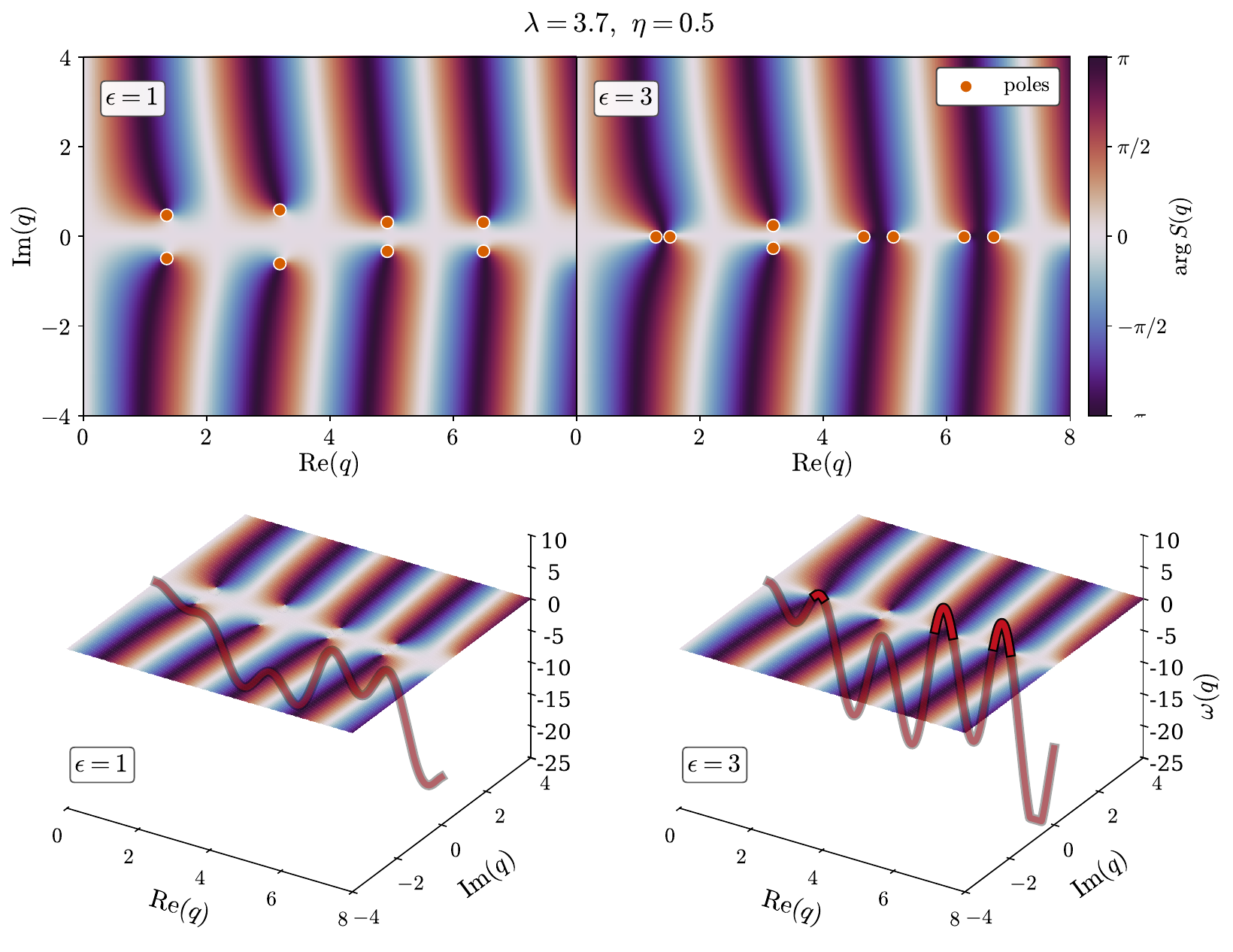}

\caption{Top: Location of the poles of the complex-valued structure factor $S(q)$ defined in Eq.~(\ref{eq:S}) in the complex plane $q=k_r+ik_i$ for a system with $\lambda = 3.7$, $\eta = 0.5$ and two different temperatures, $\epsilon = 1.0$ (left panels) and $\epsilon = 3.0$ (right panels). The positions of the poles, where $1/S(q)\to0$, can be tracked by following the locations of the topological defects in the argument, $\arg S(q)$. Conjugate pairs of poles near the real axis give rise to peaks in the dispersion relation $\omega(k)$ along the real axis, where $q=k$ is real. In the lower plots, $\omega(k)$ is displayed as the red line. When $\omega(k)>0$, a density mode becomes unstable and the corresponding poles become real-valued; these are the zero-crossings of $\omega(k)$. Hence, the position of the maxima $k_u$ of $\omega(k)$ lying between two consecutive poles along the real axis can be identified as wavenumbers corresponding to unstable density-modes.}
\label{fig:2}
\end{figure}

Having shown that the OZ route results are rather accurate and therefore the results we have for $\hat{c}^{(2)}(k)$ in Eqs.~\eqref{eq:c_split}, \eqref{c_FMT_hc} and \eqref{phi:ss_k} are also reliable, we can now use these to determine the pole structure in the complex $q$ plane.
In Fig.~\ref{fig:2} we display examples at two different temperatures of where the poles are located, for the square-shoulder system with $\lambda=3.7$ and a (reservoir) packing fraction of $\eta=0.5$.
The two different temperatures correspond to $\epsilon=1.0$ on the left and $\epsilon=3.0$ on the right, respectively.
The red circles indicate the locations of (complex) poles and the colour of the background heatmap denotes the phase of the (complex-valued) structure factor, i.e.\ $\arg S(q)$.

The detection of poles of the complex-valued structure factor $S(q)$ in Eq.~\eqref{eq:S} is realized by tracking ``topological defects'' in the phase $\phi=\arg S(q)$.
To this end we apply a simple, yet (with sufficient resolution) robust algorithm that is used in the context of vortex line tracking in random wave patterns \cite{Taylor2016-qd}. The poles are located by evaluating $S(q)$ on a regular Cartesian grid with spacing $\Delta q=0.013/\sigma$ and then identifying $2\times2$ neighbouring grid points with consistent winding orientation.
Thinking of this as an image analysis algorithm, then this corresponds to identifying properties of $2\times2$ pixel stencils.
Thus, in the present context, we refer to {\em pixel} as a discrete lattice position in the complex plane and the $2\times 2$ pixel stencil is formed by four lattice positions on a square.
The algorithm consists of two steps: 

\begin{itemize}
\item [a)] for each boundary between pixels $i$ and $j$ compute the phase difference and orient it into the direction that minimizes the absolute phase difference modulo $2\pi$. This can be expressed as the principle argument $\triangle\phi_{ij}=\arg(e^{i(\phi_j-\phi_i)})$ where $\arg(\cdot)\in(-\pi,\pi]$ or equivalently as $\triangle\phi_{ij}=(\phi_j-\phi_i)-2\pi\lfloor\frac{\phi_j-\phi_i+\pi}{2\pi}\rfloor$. 
\item[b)] Within each $2\times 2$ pixel stencil the four orientations of phase differences must have identical orientations when traversed in a (counter-)clockwise direction for a topological defect to be present.
\end{itemize}

The sum of the four discrete phase differences approximates the closed contour integral of the winding number $W=\frac{1}{2\pi}\oint_C\nabla\phi(r)dl$, which in our case is $\pm1$. 

In the fluid phase, the poles are complex with non-zero imaginary part and for $\epsilon\geq0$ appear in pairs of complex conjugates, as can be seen on the left hand side of Fig.~\ref{fig:2}.
To determine $h(r)$ for all $r$, one must sum over contributions from all of the displayed poles, as well as the (in principle) infinite more poles in the upper half of the complex $q$ plane that are not displayed.
However, the large-$r$ asymptotics of $g(r)$ is dominated by just the leading order pole, which possesses the smallest imaginary part, because this pole corresponds to the slowest decay as given in Eq.~(\ref{hr_asym}).
Additionally, the real part of the pole determines the wavelength of oscillation in the asymptotic decay, i.e.\ determining the dominating length scale in the pair correlation.

Some additional insight into the form of the structure factor $S(q)$ in Eq.~(\ref{eq:S}) and the dispersion relation $\omega(k)$ in Eq.~(\ref{eq:omega}) can come from inspecting Fig.~\ref{fig:2}.
While $S(q)$ (and so also $\omega(k)$) is a real function on the real axis, they manifest signatures of the close complex poles in form of maxima.
In the lower panel of Fig.~\ref{fig:2} we show the structure of the dispersion relation $\omega(k)$ in relation to the complex pole structure.
Close by poles result in a local maximum in the dispersion relation.
In the fluid phase, the dispersion relation $\omega(k)$ remains negative for all $k>0$ and modes with a given wave number $k$ decay over time -- see Eq.~\eqref{eq:d_rho}.

Where the uniform liquid is linearly unstable (i.e.\ where solid phases form), we have $\omega(k)>0$ for some $k>0$\cite{archer2012solidification, Wassermair2026}.
This is illustrated in the right-hand panels of Fig.~\ref{fig:2}, where we display the pole structure (top panel) and the corresponding dispersion relation (lower panel) for a state with the same packing fraction $\eta=0.5$ and shoulder range $\lambda=3.7$ as on the left hand side of Fig.~\ref{fig:2}, but at a significantly lower temperature, corresponding to $\epsilon=3.0$.
At this state point, there exist poles on the real axis, which imply portions of the dispersion relation being positive.
This pole structure, how they are located and the corresponding dispersion relation, which contains positive regions, is of interest for finding complicated solid structures \cite{wassermair2024fingerprints, Wassermair2026} since these define the wavenumbers of modes that can grow if the uniform liquid is quenched to such a state point.

\begin{figure}
\centering
\includegraphics[width=1.0\textwidth]{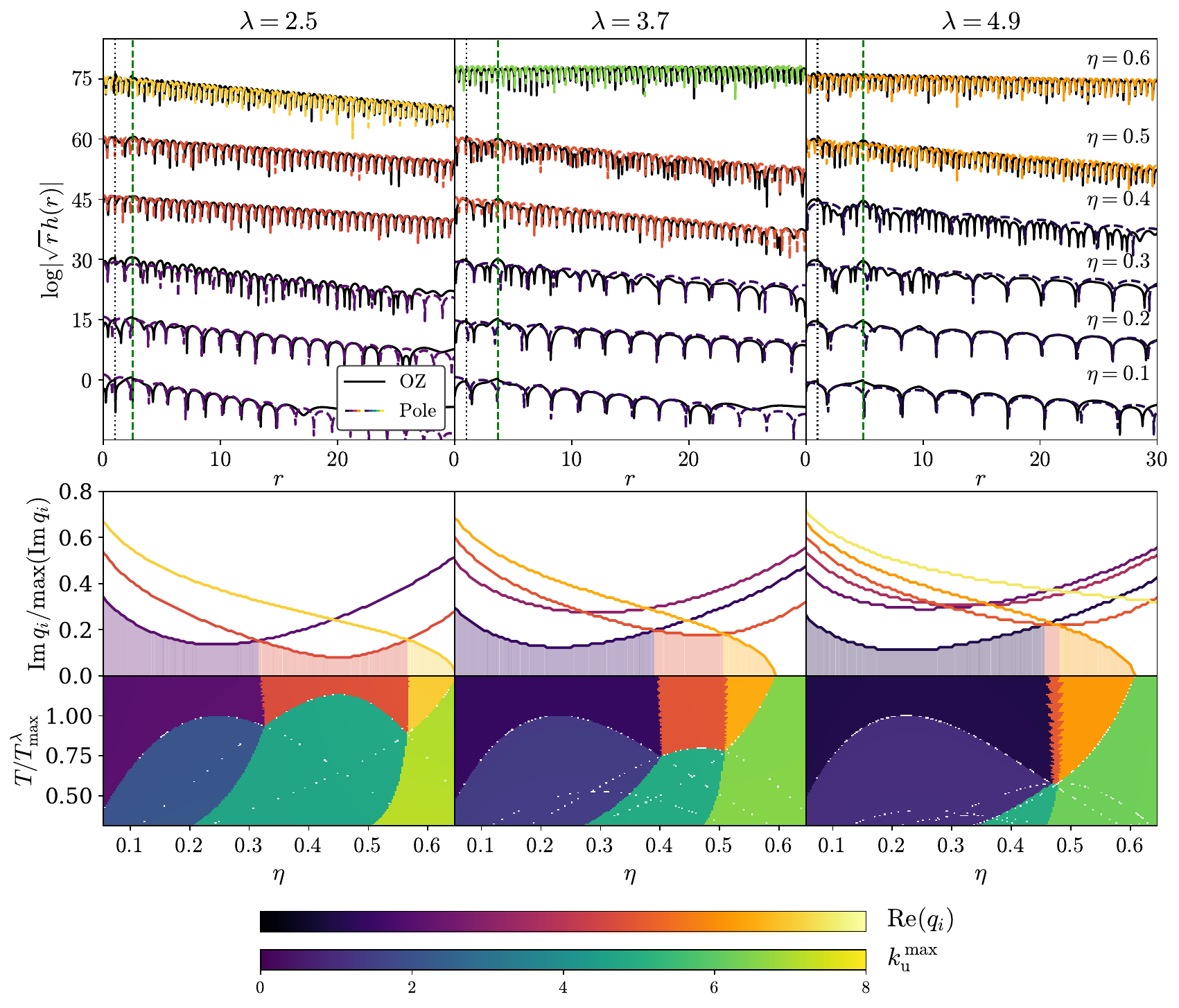}
\caption{Top panels: plots of $\log \left[\sqrt{r} h(r)\right]$, where $h(r)  = g(r) -1$, as functions of distance $r$, obtained via the OZ-route (solid black lines), compared to their asymptotic behaviour in Eq.~\eqref{hr_asym} (broken coloured lines) obtained via the pole $q_n=k_r+ik_i$ with the smallest imaginary part $\text{min}_i(\text{Im}q_i)$. These functions are calculated for three different values of shoulder range $\lambda$ (increasing from left to right, as labelled) and six different values of packing fraction $\eta$ (decreasing from top to bottom, as labelled).
For better visibility the curves are shifted vertically by 15 units. Middle panels: imaginary parts of the poles $\text{Im}q_i$ (normalized via the respective maximum values), as functions of $\eta$, with the same $\lambda$-values as in the corresponding panels above. We have also used the same temperatures as in Fig.~\ref{fig:1}, with varying packing fractions $\eta$. The shaded areas below the curves (coloured with the same colours as the respective curves) indicate the $\eta$-interval where each pole $q_i$ respectively assumes the minimum value. Lower panels: phase diagrams displaying in each case the corresponding linear stability threshold. The colouring above indicates the wavenumber of the (oscillatory) decay of $h(r)$ and the colouring below giving the value of $k_u^\mathrm{max}$, the most unstable (fastest growing) wavenumber obtained from $\omega(k)$.}
\label{fig:3}
\end{figure}

In Fig.~\ref{fig:3} we display the asymptotic decay behaviour of the
total correlation function $h(r)=g(r)-1$ for three different values of $\lambda$
and several values of the packing fraction $\eta$.
The first way
of highlighting the asymptotics is by plotting the logarithm of $h(r)$ multiplied by 
$\sqrt{r}$.
This functional form is a consequence of the leading
order expansion of the total correlation function in 
Eq.~(\ref{hr_asym}), that makes use of the pole of the structure
factor with the smallest imaginary part.
In the upper panel we
compare the full solution from
the OZ route to $h(r)$ (full black lines) with those
obtained from the leading order pole (broken coloured lines). The
packing fraction $\eta$ increases from bottom to top between
0.1 and 0.6 in increments of 0.1. For reasons of clarity
the plots for different values of $\eta$ have been shifted 
vertically. We find overall very good agreement between the two
results for both large and intermediate values of $r>\lambda$.
Some numerical problems with the OZ results at low
packing fractions and large $r$ can be seen, but the reason is easily understood by observing fast decaying correlations in those cases.

An interesting observation to be made from Fig.~\ref{fig:3} (top panels) is that the wavelength of the oscillations in $h(r)$ changes (several times in some cases, depending on $\lambda$) from a large wavelength at low packing fractions to a smaller wavelength at high packing fractions.
While such a behaviour can occur continuously, here it happens rather sharply at intermediate values of the packing fraction $\eta$.
Such a crossover has been predicted previously in a binary mixture of hard spheres (in $d=3$) based on the pole analysis of the total correlation function \cite{grodon2004decay,grodon2005homogeneous} and was called structural-crossover.
The prediction was later confirmed experimentally in effective 2D colloidal mixtures \cite{baumgartl2007experimental} and in 3D \cite{statt2016direct}.
For a one-component square-shoulder fluid in $d=1$ structural crossover was also reported \cite{Montero2025}.

In the middle panels of Fig.~\ref{fig:3} we demonstrate that the observed crossover occurs as a result of a competition between poles with different real parts, that correspond to different wavelengths.
For the system with $\lambda=2.5$ we plot the imaginary part of the three leading order poles, each having different real parts.
At low packing fraction $\eta$, the purple line corresponds to the leading order pole.
This pole has the smallest real part $\text{Re}(q)$ and so the longest wavelength oscillatory decay contribution to $h(r)$.
At $\eta\approx 0.3$ the pole corresponding to the red line becomes the leading order pole.
Note that close to the transition both poles contribute to the asymptotic decay, because both poles contribute a term with similar exponential decay.
Finally at $\eta\approx 0.55$ the pole corresponding to the yellow line displays the slowest decay.
Out of the three poles, this pole has the largest real part $\text{Re}(q)$ and so the shortest wavelength oscillatory decay contribution to $h(r)$.
For even larger values of $\eta$ the leading order pole approaches the real axes, indicating an instability of the fluid phase.

In the bottom panels of Fig.~\ref{fig:3} we sketch in each case the phase diagram as a function of $\eta$ and $T$, showing the linear stability threshold line.
In the linearly stable uniform liquid above this threshold line, the leading order pole
determines the correlation length and the main wavelength of oscillation, the value of which is given by the background colouring.
At lower temperatures, in the regions below the linear stability threshold lines, the background colouring indicates the value of $k_u^\textrm{max}$, the fastest growing wavenumber, corresponding to the largest maximum of $\omega(k)$.
This plays an important role in determining a characteristic lengthscale in the solid phases that arise in this region of the phase diagram \cite{Wassermair2026}.

As we increase the value of $\lambda$, we see in Fig.~\ref{fig:3} that the situation becomes
slightly more complicated.
For $\lambda=3.7$ we have to consider the four lowest lying poles in order to capture the behaviour of $h(r)$ in the fluid phase over the full $\eta$ range.
At sufficiently high values of $\eta$, the leading order pole again tends towards the real axes and the corresponding total correlation function $h(r)$ from the OZ route for $\eta=0.6$ (top panel) does not display a decay.
In the corresponding middle panel, we see the values of $\text{Im}q_i$ for the four leading order poles and in the bottom middle panel, the corresponding phase diagram displays the close connection to the pole structure
of the uniform fluid phase.
For the largest value of $\lambda=4.9$, considered here, we plot the imaginary parts of the six lowest poles, which allows one rationalise the oscillatory structure of the total correlation function $h(r)$ and the overall shape of the phase diagram.

\begin{figure}
\centering
\includegraphics[width=1.0\textwidth]{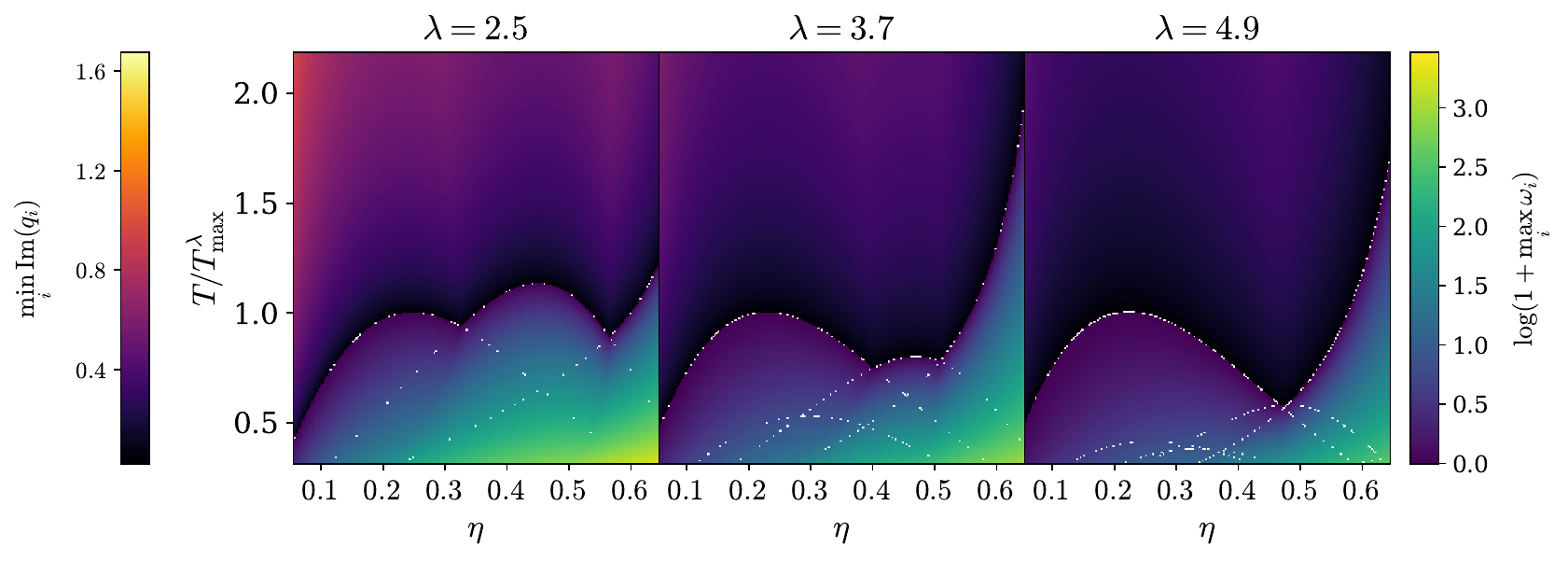}
\caption{Repeat of the phase diagrams from along the bottom of Fig.~\ref{fig:3}, showing the linear stability threshold lines for three values of $\lambda$, but here we instead colour in each case the region above the line according to the value of the reciprocal of the bulk fluid correlation length, $\min_i\text{Im}(q_i)$, and below the line with the value of $\log(1+\max_i\omega_i)$, so as the indicate the regions where the growth rate $\max\omega(k)=\max_i\omega_i$ is largest.}
\label{fig:4}
\end{figure}

In Fig.~\ref{fig:4}, we again display the phase diagram and the linear stability threshold line, but this time we colour the region above it according to the value of $\min_i\text{Im}(q_i)$, which is proportional to the reciprocal of the bulk fluid correlation length, i.e.\ the decay length of $h(r)$.
The bulk correlation length diverges on approaching the linear stability threshold, so we see that on approaching the linear stability threshold from above, $\min_i\text{Im}(q_i)\to0$.
Below the linear stability threshold, we colour the background to indicate the fastest growth rate from the dispersion relation.
The fastest growth rate is given by the maximum of $\omega(k)$, which we denote $\max_i\omega_i$.
In order to have a value on a similar scale to the quantity displayed above the linear stability threshold, we instead plot the quantity $\log(1+\max_i\omega_i)$.
We see that the growth rate increases with decreasing temperature, moving down from the linear stability threshold, and that the largest growth rates are to be found at low temperatures and higher densities, i.e.\ deep in the region where the solid phases are to be found \cite{Wassermair2026}.

\subsection{Poles for small and negative $\epsilon$}

\begin{figure}
\centering
\includegraphics[width=1.0\textwidth]{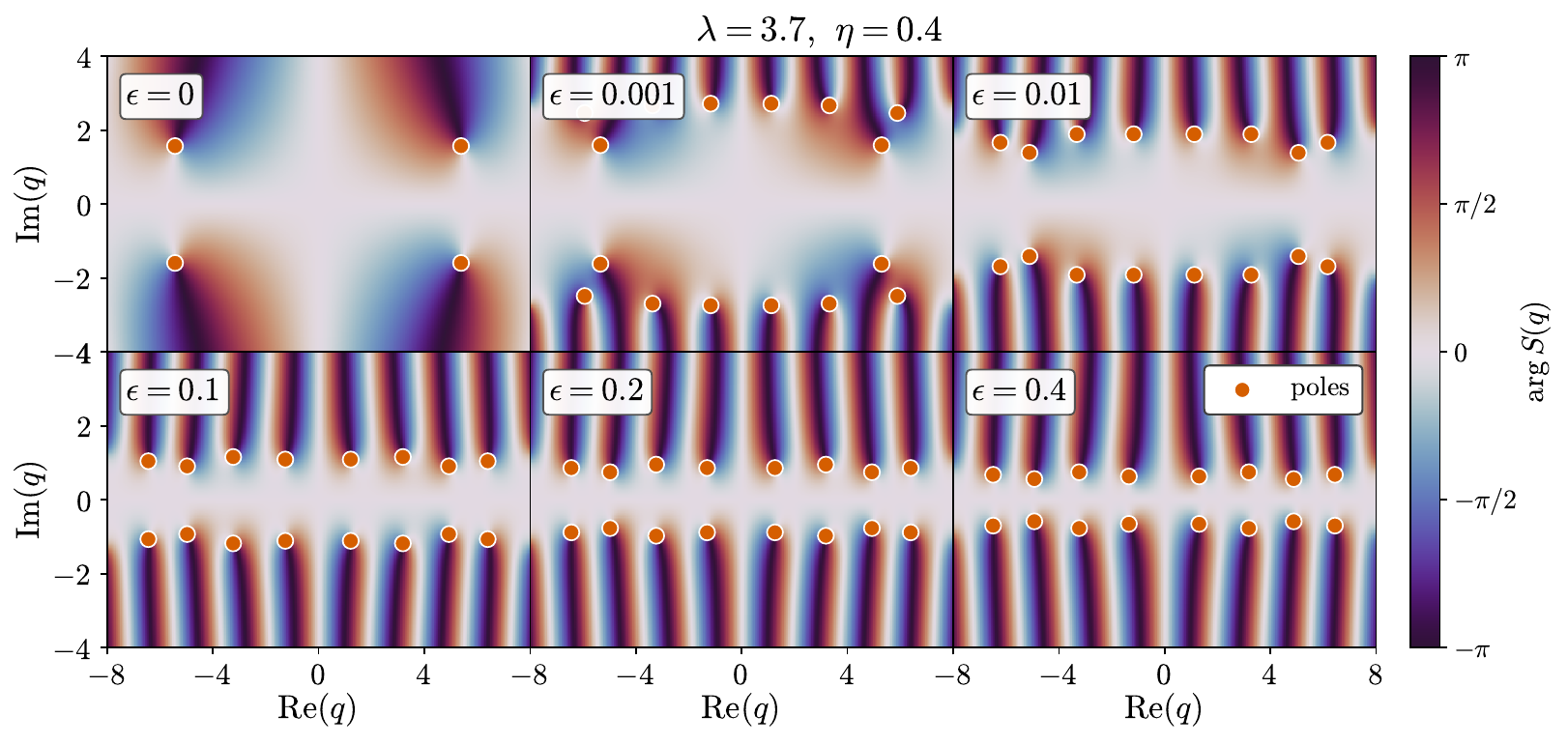}

\caption{Plots of the locations of the poles of $S(q)$ in the complex $q$-plane for $\lambda=3.7$, $\eta=0.4$ and varying $\epsilon$, as indicated above. In the limit $\epsilon\to0$, the system becomes a pure hard-disk fluid. The background colouring shows the value of the phase, $\arg S(q)$.
This should be compared with the plots at the top of Fig.~\ref{fig:2}, which are similar, but are for a higher value of $\eta=0.5$ and for the cases $\epsilon=1.0$ and 3.0.}
\label{fig:5}
\end{figure}

\begin{figure}
\centering
\includegraphics[width=1.0\textwidth]{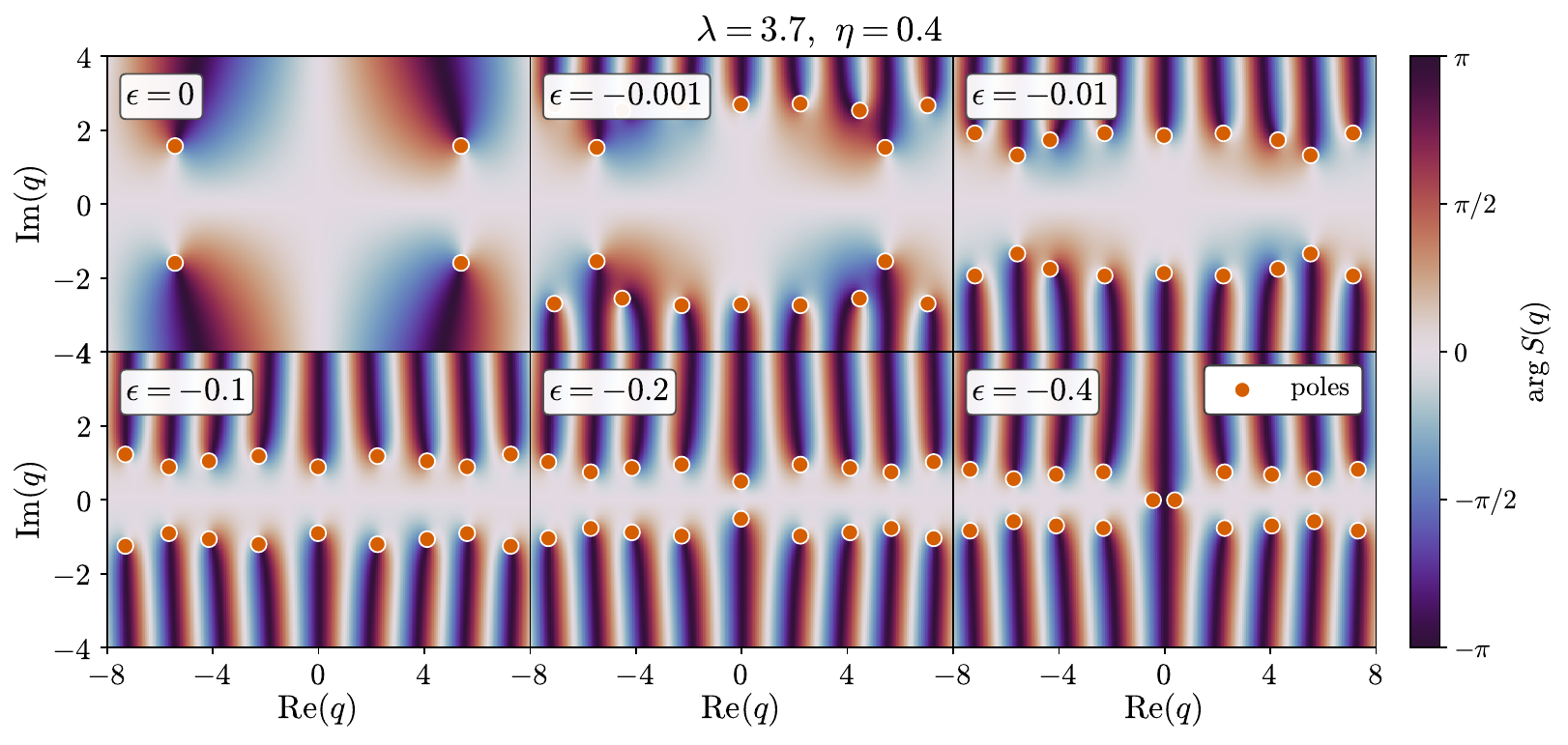}

\caption{This plot is the same as Fig.~\ref{fig:5}, except here the values of $\epsilon$ are {\em negative}, corresponding to an {\em attractive} square-well fluid.}
\label{fig:6}
\end{figure}

We now conclude our results section by briefly presenting plots showing where the poles of the structure factor $S(q)$ lie in the complex-$q$ plane (and where they move to) in two different particular cases.

The first case we consider, displayed in Fig.~\ref{fig:5}, illustrates the behaviour in the limit $\epsilon\to0$, i.e.\ when the pair potential shoulder-height tends to zero.
These results are for $\eta=0.4$, $\lambda=3.7$ and the six different values of $\epsilon=0$, 0.001, 0.01, 0.1, 0.2 and 0.4.
The value $\lambda=3.7$ is the same as for the results displayed in Fig.~\ref{fig:2}, but Fig.~\ref{fig:5} is for a somewhat lower density.
Comparing the six sets of results in Fig.~\ref{fig:5}, we go from the case $\epsilon=0$, which is just a pure hard-disk fluid, to cases with more substantial values of $\epsilon$, where the square-shoulder contribution is much more significant.
For $\epsilon$ exactly equal to zero, we have many fewer poles (only four in the portion of the complex-$q$ displayed).
For $\epsilon>0$, many more poles are introduced and we see that as $\epsilon$ is increased, these poles move towards the real axis, so that several of these poles have a comparable value of $\text{Im}(q)$ and so make a contribution to the decay of $h(r)$ and the structure of the liquid.
This illustrates how the presence of the shoulder in the pair potential introduces additional lengthscales into the liquid structure and correlations.

The second case we consider, displayed in Fig.~\ref{fig:6}, shows the behaviour when $\epsilon$ is negative, i.e.\ when the pair potential becomes {\em attractive}. Except for the sign of the chosen values of $\epsilon$, the parameter values used in Fig.~\ref{fig:6} are exactly the same as those used in Fig.~\ref{fig:5}.
At face value, the plots in Fig.~\ref{fig:6} look somewhat similar to the corresponding plots in Fig.~\ref{fig:5}, which is somewhat surprising, since an attractive shoulder is very different to a repulsive one.
On closer inspection, the main key difference is the appearance of a {\em purely imaginary} pole for negative $\epsilon$.
A purely imaginary pole leads to a monotonically decaying contribution to $h(r)$.
As $\epsilon$ becomes increasingly negative, this pole moves down the imaginary axis and eventually near $\epsilon\approx-0.3$ it hits the real axis.
This corresponds to meeting the spinodal associated with liquid-gas phase separation \cite{evans1993asymptotic}.
In contrast, for $\epsilon>0$ we see no sign of this purely imaginary pole.

\section{Discussion and Summary} \label{sec:summary}

In this study we have considered a 2D system where the particles interact via a hard core interaction plus an adjacent square-shoulder potential whose range is rather long, notably up to 4.9 times the hard core diameter.
We have calculated the RDF for the system via different routes.
Within the DFT approach we have used the highly successful FMT for the core contributions and treated the shoulder via a standard perturbation theory approach, employing the RPA.
We find that the DFT OZ route to $g(r)$ can be of comparable accuracy to the generally more reliable test-particle approach, particularly for $r>\sigma$, outside of the core.
For this reason, we have been able to apply with confidence our analytic OZ DFT results for the structure factor $S(q)$ and to determine the locations and distribution of the poles of $S(q)$ in the complex-$q$ plane.
These poles determine the asymptotic decay behaviour of the total correlation function $h(r)$ and we have found a very rich crossover behaviour in its decay as $\eta$ is varied, depending also on the particular value of $\lambda$.
Moreover, we have also been able to elucidate much about the connections between the properties of these poles and the form of the dispersion relation $\omega(k)$, which was shown to be remarkably useful for predicting properties of the crystalline and quasicrystaline phases exhibited by the present system \cite{Wassermair2026}.

Of course, the DFT we have used is not exact and so could in principle be improved upon. We are confident that the main shortcoming of our DFT approach originates from the simple perturbative approach for the soft repulsive  square-shoulder interaction, which we treat with the standard RPA ansatz, Eq.~(\ref{DFT_RPA_fex}).
This conclusion is based on the fact that for the bare hard core interaction, FMT is able to predict very accurate RDFs via the test-particle route \cite{Roth2012,Thorneywork2018}.
It should be mentioned that there are more sophisticated perturbation theories available
\cite{BH,WCA,Tschopp2020} to take into account a potential tail; however, their implementation for the system at hand is definitely beyond the scope of this manuscript.
A first step to improve the level of agreement between DFT- and simulation results, at least for the simple square-shoulder potential, might be to make use of the freedom we have to modify the shoulder interaction $\phi_\mathrm{sh}(r)$ within the hard core region, without affecting the total interaction (which is anyhow infinite inside the core). Such an approach is known as the {\em optimised} random phase approximation
\cite{Kahl1984}, 
which could improve the performance of the test-particle route.
Trying to improve the consistency between the OZ  and the test-particle route to $g(r)$ will surely be a strategy to improve the overall accuracy of the DFT. 
A step in this direction might be done either by setting $\phi_\mathrm{sh}(r)$ to zero inside the core or to replace it by a suitably designed function which is adjusted such that the agreement between the DFT and the simulation data is optimized.
To some extent, the level of self-consistency could also be improved by adjusting the core contribution of the RPA term with the 
help of exact sum rules that make use of the RDF $g(r)$ employing the test-particle route \cite{Guel2024,Guel2025}.

The work of this manuscript has even broader implications, because having an accurate theory for $c^{(2)}(r)$ is not just about getting $g(r)$ correct.
There are several other important quantities that depend on $c^{(2)}(r)$, including the isothermal compressibility.
The importance of the static structure factor and the dispersion relation is that these can be used to quickly and easily map out where in the phase diagram the solid phases possibly arise.
Recall that the dispersion relation determines the growth/decay rate of periodic density modes in the uniform liquid \cite{Archer2004} and can be 
employed to predict with remarkable accuracy the structure of the solid phases that form when the liquid becomes unstable \cite{Wassermair2026}.
In fact, by tuning the dispersion relation allows to identify pair potential parameters and state points where, for instance, quasicrystalline phases arise \cite{Wassermair2026}.
We find it surprising that the DFT used here can semi-quantitatively predict phase boundaries and the wavelengths of the density modulations that determine the crystalline and quasicrystalline structures formed by core-shoulder particle systems.
We believe all of these observations will be useful to bear in mind in future studies aimed at developing improved DFTs for core-shoulder and other systems.

\section*{Acknowledgements}

We are grateful to Florian Sanm\"uller and Matthias Schmidt for valuable comments on the manuscript and helpful discussions. The simulation results presented here were enabled via a generous allocation of CPU time by the Austrian Scientific Computing (ASC) under Project No. 71263. The authors thank Ms. Katrin Muck for her guidance related to the use of HPC. A.J.A. gratefully acknowledges support from the EPSRC under Grant No. EP/P015689/1. This research was funded in part by the Austrian Science Fund (FWF) under project no. PIN8759524 with Grant-DOI 10.55776/PIN8759524, gratefully acknowledged by GK. 

\section*{Appendix A}
\label{app_A}

We provide here further details about the OZ and test-particle routes for calculating $g(r)=1+h(r)$. The following arguments lean heavily on related arguments put forward in Ref.~\cite{Archer2017}.

\subsection{OZ equation with the RPA closure}

The exact closure relation to the OZ equation \eqref{eq:OZ} is often written as \cite{hansen13}
\begin{equation}
	c^{(2)}(r)= h(r) - \ln(h(r)+1) - \beta \phi(r) + B(r)
	\label{eq:exact _closure_relation}
\end{equation}
where $B(r)$ is termed the bridge function; $B(r)$ is in general not known exactly. The RPA-DFT approximation used here, given in Eq.~\eqref{eq:c_split} or Eq.~\eqref{phi:ss_k}, can be written as
\begin{equation}
	c_\mathrm{RPA}^{(2)}(r)= c_\mathrm{c}^{(2)}(r) -\beta \phi_\mathrm{sh}(r),
	\label{eq:c2_RPA}
\end{equation}
where $c_\mathrm{c}^{(2)}(r)$ is the pDCF for the (purely repulsive) reference hard disk fluid, given in Eq.~\eqref{eq:c2_HD}.
Plugging Eq.~\eqref{eq:c2_RPA} into the OZ equation \eqref{eq:OZ}, we obtain
\begin{equation}
	h(r)=c_\mathrm{c}^{(2)}(r) -\beta \phi_\mathrm{sh}(r)+\rho_\mathrm{b}\int\dr'h(r')c_\mathrm{c}^{(2)}(|\rr-\rr'|)-\rho_\mathrm{b}\int\dr'h(r')\beta\phi_\mathrm{sh}(|\rr-\rr'|).
	\label{eq:h_OZ}
\end{equation}
We now move on to derive a corresponding expression via the test-particle route.

\subsection{The Percus test-particle route with the RPA DFT}

Percus showed \cite{Percus1976,Vanderlick1989} that the RDF $g(r)$ is related to the density profile $\rho(\rr)=\rho(r)$ around a fixed particle  (positioned in the origin) that exerts an external potential equal to the pair potential $\phi(r)$ as: $g(r)= h(r) + 1 = \rho(r)/\rho_\mathrm{b}$. Using DFT, $\rho(r)$ may be obtained via Eq.~\eqref{eq:dOmega_drho}.
So, minimising \eqref{DFT2} with $V_\mathrm{ext}(\rr)=\phi(r)$, the resulting Euler-Lagrange equation reads
\begin{eqnarray}
	\frac{\delta\Omega[\rho]}{\delta\rho}=k_{\rm B}T
	\ln[\Lambda^2\rho(r)]
	+\frac{\delta F_\mathrm{c}[\rho]}{\delta\rho}
	+\int\dr'\rho(r')\phi_\mathrm{sh}(|\rr-\rr'|)+\phi(r)-\mu=0.
	\label{eq:EL_eq}
\end{eqnarray}
Far from the test particle, at $r\to\infty$ the density $\rho(r)\to\rho_\mathrm{b}$ and $\phi(r) = 0$.
In this limit, Eq.~\eqref{eq:EL_eq} gives
\begin{equation}
	k_{\rm B}T\ln[\Lambda^2\rho_\mathrm{b}]
	+\frac{\delta F_\mathrm{c}[\rho]}{\delta\rho}\bigg|_{\rho_\mathrm{b}}+\rho_\mathrm{b}\int\dr \phi_\mathrm{sh}(r)-\mu=0.
	\label{eq:in_bulk}
\end{equation}
If we subtract Eq.~\eqref{eq:in_bulk} from Eq.~\eqref{eq:EL_eq}, we obtain:
\begin{equation}
	0=k_{\rm B}T\ln\left(\frac{\rho(r)}{\rho_\mathrm{b}}\right)
	+\frac{\delta F_\mathrm{c}[\rho]}{\delta\rho}
	-\frac{\delta F_\mathrm{c}[\rho]}{\delta\rho}\bigg|_{\rho_\mathrm{b}}
	+\int\dr'(\rho(r')-\rho_\mathrm{b})\phi_\mathrm{sh}(|\rr-\rr'|)+\phi(r).
\end{equation}
Multiplying through by $(-\beta)$ and adding $(\rho(r)-\rho_\mathrm{b})/\rho_\mathrm{b}$ to both sides, together with making use of the following functional Taylor expansion about the bulk density:
\begin{eqnarray}
	\frac{\delta F_\mathrm{c}[\rho]}{\delta\rho}=\frac{\delta F_\mathrm{c}[\rho]}{\delta\rho}\bigg|_{\rho_\mathrm{b}}
	+\int\dr'(\rho(\rr')-\rho_\mathrm{b})\frac{\delta^2F_\mathrm{c}[\rho]}{\delta\rho(\rr)\delta\rho(\rr')}\bigg|_{\rho_\mathrm{b}}
	+H_\mathrm{c}[\rho(\rr)],
	\label{eq:Taylor}
\end{eqnarray}
where $H_\mathrm{c}[\rho(\rr)]$ denotes all higher order terms which are $\sim{\cal O}([\rho-\rho_\mathrm{b}]^2)$ and higher, we obtain:
\begin{align}
	\frac{(\rho(r)-\rho_\mathrm{b})}{\rho_\mathrm{b}}=&\frac{(\rho(r)-\rho_\mathrm{b})}{\rho_\mathrm{b}}-\ln\left(\frac{\rho(r)}{\rho_\mathrm{b}}\right)-\beta \phi(r)
	+\rho_\mathrm{b}\int\dr'\frac{(\rho(r')-\rho_\mathrm{b})}{\rho_\mathrm{b}}\left[\frac{-\beta\delta^2F_\mathrm{c}[\rho]}{\delta\rho(\rr)\delta\rho(\rr')}\bigg|_{\rho_\mathrm{b}}\right]
    \nonumber\\
	&-\beta H_\mathrm{c}[\rho(r)] 
	+\rho_\mathrm{b}\int\dr'\frac{(\rho(r')-\rho_\mathrm{b})}{\rho_\mathrm{b}}\phi_\mathrm{sh}(|\rr-\rr'|).
	\label{eq:OZ-like}
\end{align}
Now, recalling Eq.~\eqref{eq:c2}, we therefore have that
\begin{equation}
c_\mathrm{c}^{(2)}(|\rr-\rr'|)=\frac{-\beta\delta^2F_\mathrm{c}[\rho]}{\delta\rho(\rr)\delta\rho(\rr')}\bigg|_{\rho_\mathrm{b}},
\end{equation}
so that Eq.~\eqref{eq:OZ-like} becomes
\begin{align}
	h(r)=&h(r)-\ln\left(h(r)+1\right)-\beta \phi(r)-\beta H_\mathrm{c}[\rho_\mathrm{b}g(r)]
	\nonumber\\
    &+\rho_\mathrm{b}\int\dr'h(r')c_\mathrm{c}^{(2)}(|\rr-\rr'|)-\rho_\mathrm{b}\int\dr'h(r')\beta\phi_\mathrm{sh}(|\rr-\rr'|).
	\label{eq:h_DFT}
\end{align}
If $F_\mathrm{c}[\rho]$ were exact, then $-\beta H_\mathrm{c}[\rho_\mathrm{b}g(r)]$ would be the bridge-function $B_\mathrm{c}(r)$ of the reference hard disk fluid \cite{Archer2017}.
Thus, comparing the first four terms on the right hand side of Eq.~\eqref{eq:h_DFT} with the right hand side of Eq.~\eqref{eq:exact _closure_relation}, we see that these four terms together correspond to an approximation for $c^{(2)}(r)$ that is neither the exact result \eqref{eq:exact _closure_relation}, nor the RPA approximation \eqref{eq:c2_RPA}.
This close similarity to the exact expression in Eq.~\eqref{eq:exact _closure_relation}, with the only difference being the approximation effectively made for the bridge function $B(r)$, formed the basis of the arguments in Ref.~\cite{Archer2017} that mean-field DFT is often better than one might expect.
It is also one reason why the test-particle route to $g(r)$ is generally expected to be superior to the OZ route.

\section*{Appendix B}
\label{app_B}

The DFT calculations were performed using a code written by the authors that is available online via \cite{water_code}, using distributed memory parallelization of the FFT via the mpi4py-fft package \cite{jpdc_fft}.
The RDFs are obtained using Picard iteration; for details on the numerical implementation we refer to previous work using the same software \cite{wassermair2024fingerprints}. The test-particle located at $(x=0,y=0)$ is included via the shoulder part of the potential in Eq.~\eqref{eq:pot_shoulder}, treated explicitly as external potential term in the minimized free energy, while the hard core repulsion is enforced by setting the density within the core $\rho(r<\sigma) \equiv 0$, at every iteration step.
The calculations are carried out in square boxes of size $L_x=L_y=50\sigma$, discretized on $N_x=N_y=2048$ grid points using periodic boundary conditions.
The resulting density profiles are then angular-averaged according to Eq.~\eqref{eq:angav} using 500 evenly spaced radial bins.
To verify that neither the resolution nor the  boundary conditions are affecting the described features of $g(r)$, we performed DFT calculations for $\lambda=4.9$ at $\eta=0.4$ varying both box size and number of grid points and found no significant effect of both computational parameters on the resulting $g(r)$.
A calculation was considered converged, if the cumulative squared error between the densities of consecutive steps in the Picard iteration fell below $10^{-13}$.

The $g(r)$ via the OZ route are calculated from the structure factor $S(k)=[1-\rho \hat{c}^{(2)}(k)]^{-1}$ (see Eq.~\eqref{eq:c_split}) by numerical integration of the inverse Hankel transform
\begin{equation}
   g(r)=1+\frac{1}{2\pi\rho} \int_0^\infty(S(k)-1)kJ_0(kr)dk.
\end{equation}
The radius in Fourier space $k\in(0.025,250)$ was discretized uniformly with grid spacing $dk=0.025$.
The numerical integration of the above integral was then performed for the uniformly spaced real space radii $r\in\left(\frac{2\pi}{250},\frac{2\pi}{dk} = 2 \pi \cdot 40\right)$.



%

\end{document}